\documentclass{article}
\usepackage{setspace}
\usepackage[margin=1.25in]{geometry}

\usepackage{authblk}

\usepackage{graphicx}
\usepackage{subcaption}
\usepackage{caption}
\usepackage{subcaption}
\usepackage{csvsimple}
\usepackage[backend=biber,style=apa]{biblatex} %[backend=bibtex]
\usepackage{xcolor}
\usepackage{inputenc}
\usepackage{footmisc}

\bibliography{main}

\title{Measuring arousal and stress physiology on Esports, a League of Legends case study}
\author[1]{David Berga}
\author[1]{Alexande Pereda}
\author[1]{Eleonora De Filippi}
\author[1]{Arijit Nandi}
\author[2]{Eulàlia Febrer Coll}
\author[2]{Marta Reverte}
\author[3]{Lautaro J. Russo}

\affil[1]{Eurecat Centre Tecnològic}
\affil[2]{eDojo - Electronic Dojo S.L.}
\affil[3]{Mental Gaming S.L.}
\affil[*]{Corresponding author email: david.berga@eurecat.org}
\date{}
\begin{document}

\maketitle

\begin{abstract}
Esports gaming is an area in which videogame players need to cooperate and compete with each other, influencing their cognitive load, processing, stress, and social skills. Here it is unknown to which extent competitive videogame play using a desktop setting can affect the physiological responses of players' autonomic nervous system. For such, we propose a study where we have measured distinct electrodermal and cardiac activity metrics over competitive players during several League of Legends gameplay sessions in a Esports stadium. We mainly found that game performance (whether winning or losing the game) significantly affects both electrodermal and cardiac activity, where players who lost the game showed higher stress-related physiological responses, as compared to winning players. We also found that important specific in-game events such as "Killing", "Dying" or "Destroying Turret" significantly increased both electrodermal and cardiac activity over players more than other less-relevant events such as "Placing Wards" or "Destroying Turret Plates". Finally, by analyzing activity over player roles we found different trends of activity on these measurements, this could foster the exploration on human physiology with a higher set of participants in future Esports studies.
\end{abstract}

%\keywords{
%Esports, League of Legends, Competition, Physiology, Electrodermal activity, Galvanic Skin Response, Electrocardiogram, Optical Pulse, EDA, GSR, ECG, PPG, Sympathetic, Parasympathetic, Arousal, Stress, Emotion, Heart Rate.
%}

\section{Introduction}

Research in the field of psychology has traditionally focused on three main forms of both emotional and attentional responses: subjective perceptions of the individual about their own state, effects on behavior, and changes in their physiological patterns, such as acceleration or deceleration of heart rate and increase in skin conductivity \autocite{Bradley2000, Mauss2009}. Each one of these approaches shows both advantages and disadvantages. First, auto-informed methods, such as questionnaires or interviews, for which participants are directly asked to report their status are the only way to access the individual's subjective perception are limited by the individual's own ability to introspect, since many psychological processes can occur unconsciously or with low levels of consciousness \autocite{Nisbett1977}. Moreover, in certain cases it is possible that cognitive biases (such as social desirability bias) interfere with the reports, making the information not entirely reliable. For this reason, in the field of experimental psychology research, the analysis of physiological responses (for example, variations in heart rate, skin conductivity, activation of facial muscles, etc.) has been introduced as a way of obtaining information about the cognitive and emotional processes of individuals in an indirect way. On the other hand, the main disadvantage of the physiological methods with respect to the self-reported ones is that physiological experimentation requires a higher amount of resources and thus data gathering is taken in a much slower pace, so it is recommended to combine both methodologies with qualitative studies for which data can be obtained from a large number of participants, and laboratory studies with smaller samples \autocite{Cacioppo2017}. An experimental suggestion is to adapt the gathering of game events according to the type of physiological sub-measure that is desired to capture. Here are the main physiological measures that we used in our study:

\paragraph*{\textit{Electrodermal activity (or EDA, also known as galvanic skin response or GSR)}} is a correlate of the activation of the sympathetic branch of the autonomic nervous system, which provides reliable information with a high temporal resolution about participants' physiological arousal, related to the intensity of experienced cognitive-emotional responses.

\paragraph*{\textit{Cardiac activity (ECG or electrocardiogram)}, or optical pulse (PPG)} is a measure of Heart Rate. This signal is mediated by both the sympathetic and parasympathetic systems, thus responding to both physiological arousal and emotional regulation processes. This is dependent on both the intensity of emotions and their hedonic load, also with the appearance of cognitive resources for stimulus processing. The variability of heart rate can be analyzed by looking at the ratio of sympathetic vs parasympathetic activity within the PPG signal.

\subsection*{Related Work}

Physiological studies were previously carried out in the game literature \autocite{Kivikangas2011, Argasiski2017, Alhargan2017} analyzing signals such as the ECG/PPG and the GSR/EDA, showing relevant differences in the participants given their affective state, mostly validated with questionnaires. Various other studies have also compared muscle signals (Electomyography/EMG) \autocite{Ahsan2009}, brain signals (Electroencephalography/EEG) \autocite{Hafeez2021}, and facial gestures \autocite{Samara2017}. The main focus of some of these latter studies is to assess the affective state of participants from these measures, classifying each one of the 7 affective categories of basic emotions initially defined by \autocite{Ekman2005}: anger, sadness, fear, disgust, joy, surprise, contempt, and neutral. Each of the affective categories is standardized and validated with different psychophysiological tests such as the Russell test \autocite{Russell1980} or the Self Assessment Mannekin \autocite{Bradley1994}, serving as a guideline for the validation of physiological measures that could affect the affective responses of each individual. In this case, the EDA and the ECG are signals widely validated by the literature that can give insight into the changes in the affective states of each individual in a real-time manner.

In a recent review by Leis $\&$ Lautenbach \autocite{Leis2020}, 17 studies were meta-analyzed in Esports contexts for psychological and physiological stress, and it was concluded that simply playing in an Esports non-competitive environment produced no stress reactions, whereas in competitive environments several studies reported increases in anxiety levels, cortisol levels, and physiological sympathetic activation, all three indicators of stress \autocite{Jones2012, Yaribeygi2017}. However, stress is not the only interesting indicator to consider in Esports environments, whether competitive or not, peripheral physiology can also provide insight into various aspects of cognitive/emotional information processing, such as emotion, valence (polarity), engagement, boredom, frustration, etc. The benefits of extracting this information have been evident for the case of stress \autocite{Blom2019} and fatigue \autocite{Smerdov2020, smerdov2020collection}. They performed a very comprehensive sensory analysis that results in the submission of a dataset collected from 10 players on professional and amateur teams in 22 League of Legends video matches, including the simultaneous recording of the physiological (i.e. movements, pulse, saccades) and psychological (self-reported post-match survey) activity of five players, with interesting results such as the lack of correlation between stress and concentration levels for professional players. We take a similar approach, focusing on a simultaneous exploration of electrodermal activity and cardiac activity in all five players of a team in different events.
%\textcolor{brown}{(comment R2) The methodologies are introduced before the research question (or topic of research), which means there is no context to apply the chosen methodology. For example, “Related work” section is introduced before there is any context to relate the work to. There is no mention of this being a game study. My main concern in the introduction is the lack of a description/background in game studies, and specifically esports. Significant literature is missing from the introduction, and these concepts (video games and esports) are not sufficiently introduced. The paper is not at all positioned in the field, and it is not clear why this is important to explore, what impact the work will have on the field, or how it relates to the previously discussed literature in the field of psychology.}

\paragraph*{Contributions} In our study we have done the physiological analysis in an Esports stadium environment where users engage in a widely played desktop videogame "League of Legends". These are our main objectives:
\begin{itemize}
    \item Evaluate the affective responses by analyzing distinct EDA and ECG metrics depending on game performance (winning or losing).
    \item Analyse differences in physiological responses over game events (e.g. "Killing", "Dying", "Kill Assist", "Destroy Turret", "Destroy Turret Plate", "Placing Ward", etc.)
    \item Investigate statistics in individual physiological responses based on events and player roles (Jungle, Middle, Utility, Bottom, and Top)
    \item Provide the software tools necessary for the analysis and synchronization of League of Legends game events with physiological data of EDA and ECG.
\end{itemize}
%\textcolor{brown}{(comments R2: It is unclear why esports players were selected and what impact these findings have on the field of game studies and psychology. Furthermore, the main contributions highlighted in the paper (“the potential of using physiological measurements”) have already been well-established.}

\section{Methods}

\subsection{Experimental Design}

\paragraph*{Sensors and Software}

For our experiments we used a Shimmer3\footnote{\url{https://shimmersensing.com/}} pack with 5 simultaneous GSR+ Units providing Galvanic Skin Response for acquisition of Electrodermal Activity (EDA), as well as Optical Pulse (PPG) estimating heart rate variations. We developed our own Python tools (available in github\footnote{\url{https://github.com/dberga/riotwatcher-shimmer-pynput}}) for capturing data and sending it through Bluetooth (using pyserial and pylsl) and synchronizing that data with events of the game (using riotwatcher) for later statistical analysis for EDA (Ledalab\footnote{\url{http://www.ledalab.de/}} V3.4.9) and PPG (HeartPy\footnote{\url{https://github.com/smritisridhar41/Heartrate_Analysis/}}).

\paragraph*{League of Legends Skills and Mechanics}

League of Legends\footnote{\url{https://www.leagueoflegends.com/}} is a Multiplayer Online Battle Arena (MOBA) videogame developed by Riot Games \footnote{\url{https://www.riotgames.com/en}} in 2009. It has become one (if not) the most popular game of the current generation of online videogames\footnote{\url{https://medium.com/codex/the-growing-trend-of-esports-why-2023-is-going-to-be-a-massive-year-for-esports-e17fe39ac58a}} and has a significant impact on the economy of many other industrial and technology sectors (i.e. branding, consumables, social networks and media content). The original League of Legends game type mechanics (Summoner Rift) consists on a competition between 2 teams of 5 players in which each team has to destroy the enemy base or "Nexus". The battle zone is composed by 3 lanes (top, mid and bot) as well as a jungle (in between these lanes). Each team lane has 4 turrets and an inhibitor, and two more turrets in the Nexus. However, before reaching the enemy base, each team needs to destroy the turrets and the inhibitor of one of the enemy lanes. The usual teamwork consists on two players (Carry/Bottom and Utility/Support) controlling the bot lane, the Top/Tank controlling the top lane, the Middle controlling the mid lane and the Jungle that moves around all the lanes. As in other MOBAs, players will need to cooperate and sometimes play aggressively to kill the opponent's champions and overcome the enemy positions. Riot users (summoners) have a specific experience level\footnote{\url{https://leagueoflegends.fandom.com/wiki/Experience_(summoner)}} (given its in-game time) and rank\footnote{\url{https://leagueoflegends.fandom.com/wiki/Rank_(League_of_Legends)}} (given its actual performance in ranked games).

\paragraph*{Game Sessions and Subjects} A total of 4 sessions have been performed in an Asobu eSports Experience\footnote{\url{https://asobuesports.com/}} with 16 participants (contacted and selected by United Gamers Academy\footnote{\url{https://unitedgamers.pro/}}) in the gameplay experimentation. The participants were recruited with questionnaires reporting age, gender, skill level, rank level and game preferences. Only players with higher rank levels above silver III were selected for the study. Subjects' age was between 18 and 25 y.o. and 4 reported to be female whilst 12 male. Average player level was 216 (with lowest 82 and biggest 402) and corresponded to silver-gold S12 competitive rank gamers. From the 16 subjects, we captured data from 4 sessions of a total of 12 participants playing a specific team during Summoner's Rift gameplay (avg time 30-45 min), later filtered on 7 with enough valid events for statistical comparison. We cut the recording of these participants from the start to the end of the game and we set specific window times for each event (i.e. 5 sec). This data is synchronized with events downloaded from riotwatcher api\footnote{Riot API tokens available in \url{https://developer.riotgames.com/}}. Some events available for capture are "killing", "dying", "kill assist", "special killing", "item purchased", "level up", "ward placed", "building kill", "champion transform", "turret plate destroyed" and "elite monster kill". After gameplay we annotated riot's metadata for each participant such as game session data (total kills/deaths, damage done, etc.), win or loss condition and player roles (top, mid, bot, utility and jungle).

%\textcolor{brown}{(comment R2) A more detailed description of the sample and recruitment process would be helpful for study replication and future meta-analyses.}

\subsection{Physiological data processing}
\paragraph*{GSR preprocessing}

The MatLab-based toolbox Ledalab \autocite{benedek2010continuous} was used for the GSR signal preprocessing and analysis. First, we carried out a preliminary visual examination to look for periodic drift in the signal, which reflects artifacts, and  we resampled the raw signal to 50Hz using Neurokit2\footnote{\url{https://neuropsychology.github.io/NeuroKit/}}. The following preprocessing operations were then carried out using Ledalab toolbox: low-pass Butterworth filtering with a cutoff frequency of 5 Hz, and smoothing to eliminate any remaining artifacts.
Finally, we performed an event-related analysis utilizing the Continuous Decomposition Analysis to extract the features indicating Skin Conductance Responses (SCRs) (CDA).
By extracting the phasic (driver) information underlying EDA, this approach attempts to obtain the signal features of the underlying sudomotor nerve activity. Skin conductance data is deconvoluted by the overall response shape, considerably enhancing temporal accuracy.
This method enables the extraction of continuous phasic and tonic activity based on traditional deconvolution within a predetermined time window, which for us corresponded to a window comprising the three seconds before an event marker to the five following seconds. The number of SCRs within the response window, response latency for the first SCR, mean SCR amplitudes, maximum phasic, and average tonic activity within the specified window were therefore collected for each event described in the previous section. 

\paragraph*{PPG data processing}

Processing and analysis of raw PPG data were conducted using the Python-based toolkit "Heartpy" \autocite{van2019analysing, van2019heartpy}, specialized for the analysis of PPG signal as compared to ECG. At every heartbeat, blood perfuses via the capillaries and arteries, causing the skin to become discolored. The PPG detects this discoloration. The systolic peak, diastolic notch, and diastolic peak make up the signal. First, as we did with the GSR signal, we resampled the raw PPG signal to 50Hz using Neurokit2. Then, we run the processing algorithm that comes with the Heartpy toolkit and which allows for the peak detection to extract reliable time-domain measures, such as beats per minute (BPM), and Interbeat Intervals (IBI). Furthermore, for each event, we extracted measures that reflect Heart Rate Variability (HRV) such as the RMSSD (root mean square of successive differences) and the SDSD (standard deviation of successive differences). 

\section{Results}

We performed data curation for our statistical analysis using data from 7 participants (a total of 2 game sessions with recorded measures in which 3 participants played twice) with enough event samples for later analysis and processing. Some of the data not mentioned in participants results was discarded for the final evaluation due to the incorrect samples from the sensor data and/or given the lack of relevant events in the gameplay to sync with the sensor data (e.g. given too much time being dead without interacting with game objects nor players). Here we select the players' sensor data collected with enough samples to make valid statistical evaluations.
%\textcolor{brown}{(comment R2) I am unclear why the analysis was only run on 7 participants.}

%\textcolor{orange}{(comment R1) more details are needed about the statistical tests and which data is being used for each of the results.} % As a non-expert in statistics, the information provided about how and why the Friedman test was used is not sufficient which leading to doubting the results. I have some doubts that the Friedman test is applicable here but a statistician should weigh in on this. In some cases, when comparing 2 groups (winners vs losers) another test might be more appropriate like the Wilcoxon test or Mann Whitney U test.

\subsection{Physiological results: Skin Conductance}

We have processed raw GSR data with Ledalab to extract the following measures: nrSCRs (total skin conductance number "\#" of responses above threshold), Latency (delay/surpassed time "s" to elicit EDA with respect to the event), Amplitude (mean activity "mV" inside the event window), PhasicMax (max phasic value "mV" from the gap with respect the response and the event window) and Tonic (max tonic activity "mV" with respect window). Previous literature in electrodermal physiology has shown EDA can be a reliable quantifier of sympathetic dynamics \autocite{PosadaQuintero2016}, meaning higher EDA correlated with higher sympathetic (stress/alert) levels.

\begin{table}[h!]
    \centering
    %\footnotesize
    \resizebox{\textwidth}{!}{
    \begin{tabular}{|c|ccccc|}\hline
         wincon & nrSCR & Latency & Amplitude & PhasicMax & Tonic\\\hline
        WIN & 1.853±1.730 & *-.151±1.926 & *.274±.631 & .545±1.165 & *12.577±7.047\\
        LOSS & 2.599±2.210 & -.758±1.698 & .190±.414 & .370±.545 & 6.881±4.743\\
        TOTAL & 2.372±2.101 & -.574±1.788 & .215±.490 & .423±.788 & 8.611±6.121\\\hline
    \end{tabular}
    }
    \caption{Win and Loss mean GSR statistics (2 sessions) by stacking all events in one statistic. *$p<0.05$}
    \label{tab:stats-gsr-wins-losses}
\end{table}

\begin{table}[h!]
    \centering
    \resizebox{\textwidth}{!}{
    %\footnotesize
    \begin{tabular}{|c|ccccc|c|}\hline
         EVENT & nrSCR & Latency & Amplitude & PhasicMax & Tonic & N\\\hline
KILL & *1.225±1.170 & 0.0485±2.114 & *0.031±0.0416 & *0.101±0.066 & 7.764±3.233 & 9\\
DIE & 2.423±2.102 & -0.191±0.739 & 0.323±0.627 & 0.504±0.862 & *12.198±5.818 & 35\\
PLACE WARD & 2.32±2.160 & -0.295±2.145 & 0.225±0.448 & 0.152±0.138 & *12.064±5.410 & 75\\
DES.TURRET & *2.182±2.214 & -0.611±1.537 & *0.156±0.233 & 0.250±0.439 & 9.745±6.270 & 79\\
DES.PLATE & *2.175±1.885 & -0.457±1.837 & 0.294±0.587 & 0.457±0.680 & 3.087±1.768 & 49\\\hline
    \end{tabular}
    }
    \caption{Event mean GSR statistics (2 sessions) from events "Killing", "Dying", "Placing Ward", "Destroying Turret" and "Destroying Turret Plate". *$p<0.05$ }
    \label{tab:stats-gsr-events}
\end{table}
%\textcolor{orange}{(comment R1) Shouldn't the events be the same number as TABLE V?}

\begin{table}[h!]
    \centering
    \resizebox{\textwidth}{!}{
    %\footnotesize
    %\begin{minipage}{\linewidth} %0.325
    %\hspace*{0.5in}
    \centering
    \begin{tabular}{|c|c|ccccc|c|}\hline
        Event & Role & nrSCR & Latency & Amplitude & PhasicMax & Tonic & N\\\hline
KILL & Jungle & 1.200±2.168 & .644±1.578 & .039±.056 & .114±.085 & 7.974±3.427 & 5\\
& Middle & 1.333±.577 & -.713±3.271 & .024±.017 & .100±.014 & 5.993±.479 & 3\\
& Utility & .000±.000 & .000±.000 & .000±.000 & .007±.000 & 11.78±.000 & 1\\
& Bottom & .000±.000 & .000±.000 & .000±.000 & .000±.000 & .000±.000 & 0\\
& Top & .000±.000 & .000±.000 & .000±.000 & .000±.000 & .000±.000 & 0\\
\hline
DIE & Jungle & 3.000±2.793 & -.436±2.147 & .522±1.037 & .694±1.490 & 11.97±4.11 & 11\\
& Middle & .800±1.789 & -.276±.617 & .034±.076 & .163±.148 & 6.157±.780 & 5\\
& Utility & 2.333±.577 & -2.060±1.45 & .083±.039 & .041±.014 & 11.79±1.45 & 3\\
& Bottom & 2.909±2.071 & -.969±1.125 & .158±.170 & .279±.223 & 6.329±.415 & 11\\
& Top & 1.737±1.968 & -.645±1.377 & .127±.267 & .292±.333 & 10.84±9.42 & 19\\\hline
PLACE & Jungle & 3.000±2.512 & .201±2.827 & .151±.189 & .241±.250 & 9.458±4.04 & 14\\
WARD & Middle & .750±1.035 & -.598±1.109 & .029±.054 & .064±.105 & 9.40±2.91 & 8\\
& Utility & .250±.500 & .560±1.120 & .005±.010 & .238±.203 & 9.537±1.67 & 4\\
& Bottom & .000±.000 & .000±.000 & .000±.000 & .000±.000 & .000±.000 & 0\\
& Top & 2.778±.833 & -.544±1.462 & .675±1.048 & 1.335±2.00 & 21.59±2.16 & 9\\\hline
DES. & Jungle & 3.353±2.396 & -.813±1.615 & .186±.198 & .311±.342 & 14.05±3.70 & 17\\
TURRET & Middle & 1.800±2.168 & -.592±.849 & .068±.088 & .278±.214 & 5.802±.444 & 5\\
& Utility & 2.458±1.911 & -.419±2.134 & .177±.287 & .623±1.294 & 3.355±4.192 & 24\\
& Bottom & 3.077±1.935 & -.592±2.155 & .350±.562 & .546±.718 & 6.248±.617 & 13\\
& Top & 1.750±1.915 & -.488±1.661 & .082±.127 & .454±.494 & 6.731±7.723 & 16\\\hline
DES. & Jungle & 3.647±1.967 & -.673±1.915 & .678±1.02 & .844±1.136 & 11.036±2.89 & 17\\
PLATE & Middle & 1.400±1.174 & -.794±1.194 & .044±.053 & .218±.270 & 5.962±.420 & 10\\
& Utility & 2.571±2.181 & -1.298±1.51 & .148±.185 & .354±.303 & 5.897±3.913 & 21\\
& Bottom & 3.615±1.758 & -.740±2.099 & .152±.185 & .233±.198 & 6.215±.366 & 13\\
& Top & 1.278±1.965 & -.243±1.888 & .195±.489 & .368±.439 & 8.588±9.302 & 18\\\hline
ALL & Jungle & 3.125±2.380 & -.375±2.089 & .355±.716 & .488±.893 & 11.41±4.02 & 64\\
EVENTS & Middle & 1.194±1.376 & -.619±1.267 & .040±.059 & .167±.199 & 6.858±2.119 & 31\\
& Utility & 2.283±1.984 & -.778±1.861 & .144±.229 & .443±.902 & 5.465±4.515 & 53\\
& Bottom & 3.216±1.888 & -.756±1.842 & .223±.366 & .357±.468 & 6.260±.471 & 37\\
& Top & 1.758±1.853 & -.473±1.593 & .215±.523 & .507±.891 & 10.69±9.41 & 62\\\hline
    \end{tabular}
    }
    %\end{minipage}
    %\hspace*{-1.0in}
    \caption{Role-dependent mean nrSCR (\#), Latency (sec), Amplitude (mV), PhasicMax (mV) and Tonic (mV) statistics (2 sessions) from events "Killing", "Dying", "Placing Ward", "Destroying Turret" and "Destroying Turret Plate". N are event occurrences. *$p<0.05$} %(role-dependent data) are not discussed at all, they should be removed.
    \label{tab:stats-gsr-roles}
\end{table}

In Table \ref{tab:stats-gsr-wins-losses} we show mean statistics of nrSCR, Latency, Amplitude, PhasicMax, and Tonic values of players that win the gameplay and lose the gameplay. Similarly, in TABLE \ref{tab:stats-gsr-events} we show statistics for events "Killing", "Dying", "Place Ward", "Destroy Turret" and "Destroy Turret Plate". We expand these statistics in Table \ref{tab:stats-gsr-roles} filtering player roles in the game.

Given the Chi-squared measured distributions (non-parametric) we performed Friedman's tests over win-loss and event data for each GSR metric. On analyzing winning or losing the match (Tables \ref{tab:stats-gsr-wins-losses}-\ref{tab:stats-gsr-events}), we found significant differences in player's activity during "Killing" in nrSCR, Amplitude, and PhasicMax activity ($p$=.046, $\chi^2$=4.000). We also observed significant differences when winning/losing the game during "Destroying Turret" in nrSCR and Amplitude ($p$=.020, $\chi^2$=5.444) as well as "Destroying Plate" for nrSCR ($p$=.008, $\chi^2$=7.143) with a trend in Amplitude ($p$=.071, $\chi^2$=3.266). The tonic activity was only significantly distinct depending on  winning/losing for the events of "Dying" ($p$=.035, $\chi^2$=4.455) and "Placing Ward" ($p$=.002, $\chi^2$=10.000). 

We also analyzed in distributions of GSR activity between all events for the same winners and found significant differences in Latency ($p$=.024, $\chi^2$=11.265), Amplitude ($p$=.041, $\chi^2$=9.959), Tonic activity ($p$=.010,$\chi^2$=13.28) and a trend for PhasicMax ($p$=.092, $\chi^2$=8.000). In the same analysis, we did not find differences in GSR activity between the events in the case of losing the game.

\subsection{Physiological results: Heart Rate}

We have processed raw PPG data with Heartpy to obtain the BPM ("\#" beats per minute), IBI (time "ms" of interbeat interval or R-R), SDNN (standard deviation of intervals "ms" between adjacent beats of the IBI of normal sinus beats), SDSD (standard deviation of successive differences between adjacent R-R intervals "ms") and RMSSD (root mean square of successive differences between adjacent R-R intervals "ms"). The latter metrics (SDSD and RMSSD) are related to the measurement of HRV (heart rate variability). Indeed, higher HRV (higher values of SDSD or RMSSD) can represent parasympathetic/vagal activity ( associated with a state of relaxation), while a lower HRV (lower values for SDSD or RMSSD) represents sympathetic/flight-or-fight activity (being stressed or alert; \autocite{Valenza2018}). Here we have to point out that studies on HRV are commonly analyzed over large timeline streams of heart rate data (about 5 min or more; \autocite{Shaffer2017}) for more simple and large tasks. However, our measurements of HRV are considering 5 to 10-second windows of activity according to the League of Legends fast-paced events. In TABLES \ref{tab:stats-ppg-wins-losses}-\ref{tab:stats-ppg-roles} we show mean statistics of pulse metrics according to win condition, event, and role.

\begin{table}[h!]
    \centering
    \resizebox{\textwidth}{!}{
    %\footnotesize
    \begin{tabular}{|c|ccccc|}\hline
         wincon & BPM & IBI & SDNN & SDSD & RMSSD\\\hline
WIN & 172.145±112.795 & 531.156±332.714 & 128.673±48.392 & *100.174±46.513 & 201.980±87.488\\
LOSS & *93.540±50.271 & *743.045±212.297 & 76.840±52.382 & 57.060±44.479 & 117.470±92.097\\
TOTAL & 114.062±79.606 & 687.724±265.416 & 90.373±56.104 & 68.316±48.750 & 139.534±98.038\\\hline
    \end{tabular}
    }
    \caption{Win and Loss mean PPG statistics (2 sessions) by stacking all events in one statistic. *$p<0.05$}
    \label{tab:stats-ppg-wins-losses}
\end{table}

\begin{table}[h!]
    \centering
    \resizebox{\textwidth}{!}{
    %\footnotesize
    \begin{tabular}{|c|ccccc|c|}\hline
         EVENT & BPM & IBI & SDNN & SDSD & RMSSD & N\\\hline
KILL & 68.620±10.396 & 892.429±139.157 & 73.404±46.177 & 56.974±31.484 & 99.246±53.674 & 7\\
DIE & 103.381±65.949 & 725.839±249.362 & 86.549±53.718 & *67.178±46.320 & *132.395±85.887 & 37\\
PLACE WARD & 121.905±75.396 & 647.536±278.835 & 92.073±53.922 & *75.722±54.092 & 143.491±93.544 & 31\\
DES.TURRET & 118.132±94.834 & 678.146±260.378 & 94.580±55.782 & 72.386±51.682 & 145.934±96.149 & 57\\
DES.PLATE & 117.418±78.046 & 672.916±275.895 & *89.918±60.236 & *63.527±46.951 & 140.361±110.991 & 71\\\hline
    \end{tabular}
    }
    \caption{Event mean PPG statistics (2 sessions) from events "Killing", "Dying", "Placing Ward", "Destroying Turret" and "Destroying Turret Plate". *$p<0.05$}
    \label{tab:stats-ppg-events}
\end{table}

After performing Friedman tests over PPG metrics for all events and we found significant differences in SDSD ($p=$.016, $\chi$=10.371) when winning the game, while when losing the game we found differences in BPM and IBI ($p=$.041, $\chi$=8.28). We also tested if there were differences when winning or losing the game for each specific event and we saw significant differences for "Destroying Turret Plate" between win and loss for SDSD ($p=$5.32$\times10^{-4}$, $\chi$=12.0) and RMSSD ($p=$.004, $\chi$=8.333), and "Dying" (SDSD; $p=$.011, $\chi$=6.4)(RMSSD; $p=$.002, $\chi$=10) as well as for SDSD when "Placing a Ward" ($p=$.011, $\chi$=6.4).

\begin{table}[h!]
    \centering
    \resizebox{\textwidth}{!}{
    %\footnotesize
    %\hspace*{-1.0in}
    \begin{tabular}{|c|c|ccccc|c|}\hline
    Event & Role & BPM & IBI & SDNN & SDSD & RMSSD & N\\\hline
KILL & Jungle & 63.195±10.832 & 967.750±142.479 & 84.5987±57.767 & 66.8081±40.275 & 113.310±67.023 & 4\\
& Middle & 75.852±3.301 & 792.000±34.176 & 58.4780±28.399 & 43.8612±9.449 & 80.492±31.328 & 3\\
& Utility & .000±.000 & .000±.000 & .000±.000 & .000±.000 & .000±.000 & 0\\
& Bottom & .000±.000 & .000±.000 & .000±.000 & .000±.000 & .000±.000 & 0\\
& Top & .000±.000 & .000±.000 & .000±.000 & .000±.000 & .000±.000 & 0\\\hline
DIE & Jungle & 75.215±8.200 & 806.800±93.180 & 69.730±48.760 & 49.185±41.834 & 99.139±63.691 & 10\\
& Middle & 80.846±8.515 & 748.667±77.577 & 67.150±31.274 & 55.873±25.619 & 95.511±31.233 & 5\\
& Utility & 176.58±60.409 & 375.714±157.912 & 158.366±16.102 & 104.249±26.433 & 254.481±58.683 & 3\\
& Bottom & 71.112±5.453 & 848.111±64.272 & 58.362±25.412 & 48.458±29.071 & 97.162±44.958 & 9\\
& Top & 149.90±100.33 & 628.457±392.726 & 116.888±63.553 & 96.551±58.618 & 179.177±108.76 & 37\\\hline
PLACE & Jungle & 72.025±8.956 & 845.821±111.874 & 68.544±40.886 & 54.092±32.402 & 104.573±69.816 & 13\\
WARD & Middle & 75.660±5.498 & 796.250±59.320 & 60.883±25.517 & 41.489±21.775 & 80.912±39.805 & 4\\
& Utility & 151.11±70.545 & 513.417±302.925 & 111.647±53.012 & 96.855±71.343 & 175.363±99.071 & 8\\
& Bottom & .000±.000 & .000±.000 & .000±.000 & .000±.000 & .000±.000 & 0\\
& Top & 221.87±73.442 & 297.601±101.160 & 137.749±61.433 & 117.229±51.861 & 227.036±95.086 & 6\\\hline
DES. & Jungle & 81.985±10.532 & 742.917±92.725 & 66.538±35.728 & 46.652±31.067 & 85.828±44.360 & 15\\
TURRET & Middle & 96.998±49.985 & 696.306±174.643 & 83.246±45.988 & 53.699±26.111 & 128.027±86.284 & 12\\
& Utility & 176.16±91.996 & 455.294±261.896 & 144.335±49.317 & 121.310±43.578 & 241.965±84.882 & 14\\
& Bottom & 86.635±19.452 & 721.889±157.714 & 127.253±59.682 & 110.750±68.850 & 184.948±96.020 & 6\\
& Top & 135.37±176.90 & 844.944±376.067 & 60.987±44.726 & 41.902±40.029 & 99.731±75.180 & 10\\\hline
DES. & Jungle & 76.362±8.247 & 795.708±100.562 & 60.503±46.094 & 46.599±35.956 & 87.578±64.595 & 16\\
PLATE & Middle & 76.874±6.404 & 785.385±63.742 & 61.381±29.526 & 47.089±34.529 & 87.474±37.099 & 13\\
& Utility & 202.95±88.446 & 355.705±182.813 & 129.698±42.153 & 111.779±44.273 & 205.033±81.351 & 16\\
& Bottom & 96.659±35.939 & 691.758±214.864 & 113.291±80.127 & 51.086±25.804 & 193.149±167.07 & 11\\
& Top & 120.34±94.250 & 769.005±378.607 & 86.455±68.067 & 53.487±50.697 & 134.803±127.02 & 15\\\hline
ALL & Jungle & 75.739±10.170 & 807.065±114.546 & 67.119±42.557 & 50.132±34.495 & 94.703±59.744 & 58\\
EVENTS & Middle & 83.723±29.622 & 753.243±116.137 & 68.963±35.358 & 49.553±27.323 & 100.437±58.497 & 37\\
& Utility & 181.76±84.367 & 421.948±236.871 & 133.272±46.371 & 111.570±48.563 & 215.473±86.246 & 41\\
& Bottom & 85.503±27.045 & 752.833±172.923 & 97.499±66.037 & 63.945±46.580 & 158.030±125.32 & 26\\
& Top & 146.07±119.30 & 684.261±387.718 & 95.172±64.519 & 70.493±56.388 & 150.569±111.94 & 41\\\hline
    \end{tabular}
    }
    %\hspace*{-1.0in}
    \caption{Role-dependent mean BPM (\#), IBI (ms), SDNN (ms), SDSD (ms) and RMSSD (ms) statistics (2 sessions) from events "Killing", "Dying", "Placing Ward", "Destroying Turret" and "Destroying Turret Plate". N are event occurrences. *$p<0.05$} %(role-dependent data) are not discussed at all, they should be removed.
    \label{tab:stats-ppg-roles}
\end{table}

\section{Conclusions}

This study shows the potential of using physiological measurements (EDA and ECG) over Esports and desktop gameplay environments. In this study, we characterized physiological responses depending on performance, events as well as participants' roles in the game. In most cases, we found significant differences in EDA (for nrSCR, Amplitude, and PhasicMax activity) during "Killing", "Destroying Turret" or "Destroying Turret Plate" between players that are winning the game and players that are losing the game. When players were winning the game, they showed distinct patterns of physiological activity depending on the events in the game (e.g., "Killing", "Destroying Turret", "Destroying Plate"). In contrast, we did not find any significant difference between these events for players that were losing the game. This can hinder the possibility that players that perform badly show similar physiological states across the game, while players that perform well have distinct physiological behavior during the game. For the case of ECG, SDSD was significantly distinct for players that were winning the game between different events.  On the other side, we found that only IBI and BPM measures showed significant differences for players that were losing the game. Overall, players that performed better (winning) showed significantly higher parasympathetic activity (i.e., relaxation) than the ones that were losing. 

These results suggest that poor game performance induces higher stress or to be in a state of alert, while players that perform better tend to remain in more relaxed states. Furthermore, the analysis for specific events, like "Dying", "Destroying Turret Plate" or "Placing Ward", has shown that players have distinct values of SDSD and RMSSD, with Killing" or "Dying" events inducing higher sympathetic activity (lower HRV). 

Despite the lack of physiological samples for participants and game sessions we obtained enough measurements to pinpoint differences in-game performance and events. By having a higher number of participants and game sessions we would suggest undergoing similar studies, not only for analyzing physiology over game performance and events but also for conducting an in-depth analysis of game roles, champions, player level, and type of match (beyond League of Legends' summoner's rift) in relation with EDA and ECG measurements. %\textcolor{brown}{(comment R2) The conclusions are not supported by the data. A causal relationship cannot be established between performance and arousal. Furthermore, because each event was discrete in the context of a longer play session, it is unclear how momentary performance on discrete events is related to overall states of “winning/losing”, which the authors argue for.}

\subsection*{Author Contributions} 
David Berga contributed to the development, data analysis, experimentation, writing and reviewing of the paper. Alexandre Pereda contributed to the experimental design with subjects and reviewing the paper. Eleonora de Filippi contributed to the data analysis and statistics as well as writing and reviewing of the paper. Arijit Nandi contributed to the development tools of shimmer. Eulàlia Febrer, Marta Reverté and Lautaro Russo contributed to the session management and contact with parters for the experimental setting. 

\subsection*{Funding}
This study has been possible through the Grant IRC 2020 (Reference ACE033/21/000046) funded by ACCIÓ (Catalan Agency for Business Competitiveness), from the project "eSports-LAB" lead by INDESCAT (Associació Catalana Clúster de la Indústria de l'Esport), partners with Generalitat de Catalunya, EsportCat and Fundació Catalana per l'Esport.

\subsection*{Conflicts of Interest}
The authors declare that they have no known competing financial interests or personal relationships that could have appeared to influence the work reported in this paper. Experiment participants signed an authorization form prior to the study to authorize the usage of the captured physiological data as well as performance from their Riot Gamertag and remained anonymous according to the Spanish national law LOPD (Ley Orgánica de Protección de Datos de Carácter Personal).

\printbibliography

\end{document}